  \documentstyle[12pt]{article}
\newcommand{\bmth}[1]{\mbox{\boldmath{$ #1$}}}
\newcommand{\epsl}{\varepsilon}

\title{
 \begin{flushright} {\normalsize KNK 9452} \end{flushright}
 Two-Body Dirac Equation and Its WFO
 \thanks{The main part of the paper was presented in the workshop "Fundamental
 Problems in the Elementary Particle Theory" held at Nihon University in March
 1995. }
}
\author{{ 
Hitoshi I{\sc to}} \\
 \normalsize \it Department of Physics, Faculty of Science and
Technology \\ \normalsize \it Kinki University, Higashi-Osaka, 577 J{\sc apan}}
 \date{\normalsize \it November 15, 1995 }
\begin{document}

\maketitle
 \small
 {\bf Abstract}

A relativistic equation is deduced for the bound state of two particles,
by assuming a proper boundary condition for the propagation
 of the negative-energy states.
 It reduces to the (one-body)Dirac equation in the infinite limit of one of the
constituent mass. It also has the symmetries to assure the
 existence of the anti-bound-state with the same mass.
The interaction kernel(pseudo-potential) is systematically
constructed by diagonalizing the Hamiltonian of the background field theory, by
 which the retardation effects are included in the interaction.  Its wave
function at the origin(WFO) behaves
 properly in a manner sugested by the covariant field theory.

\normalsize




\section{Introduction}

One of the unsatisfactory nature of the Bethe-Salpeter equation for the bound
state is that it does not reduce to the Dirac equation in the infinite limit
of the one of the constituent mass, when the interaction is assumed to be
 instantaneous\cite{Salpeter52}. We have to sum up all the
crossed diagrams to recover the Dirac equation\cite{Brodsky69}, which is
 impossible for the finite masses.

Historically,
 the relativistic single-time equation for the two-body system preceded
 the BS equation. Soon after the discovery of the Dirac equation, Breit
 proposed the equation of the form\cite{Breit29}

\begin{equation}
  \{ H_1(\bmth{p}_1)+H_2(\bmth{p}_2)+V \}\psi=E\psi,  \label{1}
\end{equation}
where $H_i$ is the Dirac Hamiltonian

\[  H_i(\bmth{p}_i)= \bmth{\alpha}_i\cdot\bmth{p}_i + m_i\beta_i   \]
and $V$ is a local potential. The Breit equation reduces to the Dirac equation
 in the limit mentioned above but does not have the "$E$-parity symmetry", by
which we mean that there is symmetric negative eigenvalue $E$ for each positive
one, which is interpreted as the bound state of the antiparticles.

$E$-parity symmetry is a consequence of the $PCT$ invariance, for which we
assume the non-quantized transformations.
 The Dirac Hamiltonian is odd under the $PCT$ transformation and the
interaction Hamiltonian considered below is invariant under it. We therefore
see
that the $E$-parity symmetry in the instantaneous BS equation is assured
 by a projection factor
\begin{equation}
     \Lambda_{++}-\Lambda_{--}    \label{2}
\end{equation}
 in front of the potential $V$, which consists of the
 energy-projection operators

\begin{equation}
  \Lambda_{\epsl\eta}(\bmth{p}_1,\bmth{p}_2)=\Lambda_\epsl^1(\bmth{p}_1)
  \Lambda_\eta^2(\bmth{p}_2), \hspace{5mm} \epsl, \eta=+\mbox{ or }-, \label{3}
\end{equation}
where

\[ \Lambda_\epsl^i(\bmth{p}_i)=\{E_i(p_i)+\epsl H_i(\bmth{p}_i)\}/2E_i(p_i), \]
\[ E_i(p_i)=(\bmth{p}_i^2+m_i^2)^{1/2}.  \]
It is necessary to introduce a similar factor in any attempt at the
construction of an improved two-body equation.

The factor (\ref{2})
 comes from the St\"{u}ckelberg-Feynman boundary condition
 for the propagation of the negative-energy state\cite{Stuckelberg41}.
 We will construct the equation for the unequal-mass constituents by
imposing similar boundary condition and also investigate the
equal-mass equation.
 But, before presenting it we should restrict the framework of consideration.
For definiteness, we assume Abelian gauge fields interacting with the Dirac
 particles. We also work in the rest
 frame of the bound system($P=(E,\bmth{0})$),
 since we are looking for the non-covariant
approximation of the low-energy dynamics. $\bmth{p}$ and $\bmth{x}$, in the
 following, are the relative momentum and coordinate, respectively,
 in this frame.

\section{Unequal-mass equation and its properties}

By assuming a boundary condition for the propagation of the negative-
energy states, we deduce a new equation, which we call the Two-body
Dirac equation. We first consider the unequal-mass constituents and
assume that the mass $m_1$ is larger than $m_2$.

\subsection{\it Two-body Dirac equation}

We start with the pseudo-4-dimensional form of the equation in the
 momentum space

\begin{equation}
 \psi(p)=iS_F^{(2)}(P,p)\int V(\bmth{p},\bmth{q})\psi(q)d^4q/(2\pi)^4,
                                                              \label{p4}
\end{equation}
where $S_F^{(2)}(P,p)$ is the 2-body propagator in the lowest order and
 $V(\bmth{p},\bmth{q})$
represents the interaction, which is assumed not to depend on the relative
energies but is not necessarilly an instanteneous local potential.
 We impose the boundary condition that the negative-energy states
propagates backward in time. If we retain individuality of the constituents and
 use the usual Feynman propagator the factor (\ref{2}) results.
However, we can choose the other possibility in which
 we incorporate the idea that the bound two bodies should
 be treated as a quantum-mechanical unity.\footnote{We can establish
 the concept of individuality in the quantum
mechanics only through observation, which brings about a subtle point to the
bound system even in infrared-free theories:
 To detect an individual one in the bound constituents, we
need to separate them by applying the 3rd interaction, which inevitably
destroys the original state. So there is no reason why we have to apply the
free propagator individually to each constituent.}
 Since $m_1$ is larger than $m_2$,
 the free part of the Hamiltonian has the same sign as that of the particle 1
in the CM system. Let us then  modify the boundary condition as follows:
{\it A bound two-particles state propagates backward in time if
 their net energy is negative.}  By assuming it, we have the lowest propagator

\begin{equation}
  S_F^{(2)}(P,p)=\sum_{\varepsilon\eta}
\frac{1}{ \lambda_1 E-p_0-H_1(-\bmth{p})+i\varepsilon\delta}
\frac{1}{\lambda_2 E+p_0-H_2(\bmth{p})+i\varepsilon\delta}
\Lambda_{\varepsilon\eta}\gamma_0^1\gamma_0^2,    \label{2p}
\end{equation}
where $\lambda_1+\lambda_2 =1$ and the limit $\delta\rightarrow +0$ is assumed.

After integrating out the redundant degree of the freedom in (\ref{p4}), we
get the required equation

\begin{equation}
  \{ H_1(-\bmth{p})+H_2(\bmth{p})+\sum_{\epsl\eta}\epsl\Lambda_{\epsl\eta}V
\}\psi=E\psi,  \label{4}
\end{equation}
which reduces to the Dirac equation in the infinite limit of $m_1$.
 It is easy to see the $E$-parity symmetry of this equation.

If we would apply our equation to the scattering state, we shall have,
 from the time-dependent equation, the conserved probability density

\begin{equation}
   \rho(\bmth{x},t)=\psi(\bmth{x},t)^\dagger \sum_{\epsl\eta}\epsl
     \Lambda_{\epsl\eta} \psi(\bmth{x},t),   \label{5}
\end{equation}
which is in accord with the boundary condition that the negative-energy state
 propagates backward in time carrying the negative probability
density\cite{Aitchison82}.\footnote{ We note that
 we can interprete the causal propagator of a Dirac particle, applied to
one-particle wave function with the negative energy, also as carrying the
negative probability density\cite{Bjorken64}.}
But, it does not necessarily provide the normalization condition for the
bound state. For the scattering processes, the projected wave function
$\Lambda_{\epsl\eta}\psi$ corresponds to the physical state of the
(free)particles with the positive or negative energy $E$. And the above
interpretation of the probability current actually says that
{\it the state with the negative $E$ is carrying the negative
probability density}. However, for the
bound state with the positive eigenvalue $E$,
 $\Lambda_{--}\psi$ or $\Lambda_{-+}\psi$ is merely {\it a negative-energy
 component in the representation in which the energy of the free particle
 is diagonal}. It is like a small component of the one-body Dirac equation.

Taking the above consideration into account, we restore the probability
interpretation of the wave function and normalize it by assuming
 the probability density

\begin{equation}
   \rho(\bmth{x})=\psi(\bmth{x},t)^\dagger \psi(\bmth{x},t).  \label{6}
\end{equation}
Observables except for the Hamiltonian are self-adjoint under this metric:
\begin{equation}
  (\phi,\hat{O}\psi)=(\hat{O}\phi,\psi).                \label{66}
\end{equation}
The Hamiltonian is the operator ruling the time development of the system and
is modified by the factor $\sum_{\epsl\eta}\epsl\Lambda_{\epsl\eta}$. Though
it is not a self-adjoint operator, its eigenvalue is proved to be real if the
inner product (\ref{g3}) below exists.

 \subsection{\it Green's function and the vertex equation}

The Green's function $G$ for (\ref{4}) satisfies the operator equation
\begin{equation}
  \{E- H_1 -H_2 -\sum_{\epsl\eta}\epsl\Lambda_{\epsl\eta}V \}G
                     =\sum_{\epsl\eta}\epsl\Lambda_{\epsl\eta},  \label{g1}
\end{equation}
and
\begin{equation}
  G\{E -H_1 -H_2 -V\sum_{\epsl\eta}\epsl\Lambda_{\epsl\eta} \}
                   =\sum_{\epsl\eta}\epsl\Lambda_{\epsl\eta}.  \label{g2}
\end{equation}
In the momentum representation, it can be written, by using the eigen-function
$\chi_n(\bmth{p})$ of (\ref{4}), as
\begin{equation}
 G(\bmth{p},\bmth{p}')=\sum_n \frac{1}{N_n(E-M_n)}
         \chi_n(\bmth{p})\chi_n(\bmth{p}')^{\dag} +\mbox{continuum},
\end{equation}
where $M_n$ is an eigenvalue and $N_n$ is a normalization factor defined by
\begin{equation}
  N_n=(\chi_n,\sum_{\epsl\eta}\epsl\Lambda_{\epsl\eta}\chi_n).  \label{g3}
\end{equation}
When one of the constituents(labeled with 2) is in the category of the
 antiparticle of the other, there can be an annihilation process for which
the unamputated-decay-vertex $\Phi$ is given by
\begin{equation}
   \Phi= C\gamma G,  \label{g4}
\end{equation}
where $\gamma$ is the lowest vertex and $C$ is the charge-conjugation matrix
 of the particle 2.

$\Phi$ satisfies the vertex equation
\begin{equation}
 \Phi(E-H_1-H_2- V\sum_{\epsl\eta}\epsl\Lambda_{\epsl\eta})=
            C\gamma\sum_{\epsl\eta}\epsl\Lambda_{\epsl\eta}  \label{g5}
\end{equation}
and the amputated vertex is
\begin{equation}
  \Gamma =\gamma +C\Phi V.                             \label{g6}
\end{equation}
We can determine the renormalization constant $Z_1$ for the wave function at
the origin(WFO) from (\ref{g5}) and (\ref{g6}), if we need it\cite{Ito82}.

\subsection{\it Interaction Hamiltonian}

We have, so far, not specified the interaction Hamiltonian(quasipotential).
In this section, we investigate it for the one-(Abelian)guage-boson exchange
in the Coulomb guage as an example.
 For the instantaneous part of the interaction, $V$ is obvious. For the
remaining part, we can specify the quasipotential in a clear way
 from the background field theory. We have already
 shown, for the Salpeter equation, that the quasipotential from the
 one-boson exchange is given through the diagonalization of Fukuda,
 Sawada, Taketani\cite{Fukuda54} and Okubo\cite{Okubo54}(FSTO)\cite{Ito90}.
However, we first have to correct some error in
 Ref.\cite{Ito90}.\footnote{The error is only conceptual for the Salpeter
equation. The result of Ref.\cite{Ito90} is correct.}
 Namely, we employed the usual Fock space and reinterpreted the matrix elements
 of the interaction Hamiltonian including the negative-energy indices as the
 one in this space. The guiding principle was the hole theory.
But it brings the procedure into confusion,
since we have revived the negative energy in Eq.(\ref{4}).
The correct choice is to generalize the Fock space
 by ignoring the hole theory. The hole theory is partially recovered afterward
 by including the projection factor in Eq.(\ref{4}).

We show only the lowest one-boson exchange potential in the following. We first
 introduce the generalized Fock subspace of the free constituents, the bases
of which are denoted by $ |\epsl,\eta,\bmth{p}\rangle, $
where $\epsl$ and $\eta$ are the signs of the energies of the
particles 1 and 2 respectively. We then diagonalize the Hamiltonian in the
Schr\"{o}dinger picture by using the FSTO method. The second-order
boson-exchange potential in this subspace is given by

\begin{eqnarray}
\lefteqn{\langle\epsl,\eta,\bmth{p}|V(\mbox{1b})|\epsl',\eta',\bmth{p}'\rangle=
\frac{g^2}{(2\pi)^3}
\sum_{ij}\alpha_{1i}(\delta_{ij}-\frac{1}{\bmth{q}^2}q_iq_j)\alpha_{2j} }
 \nonumber
  \\ & & \times\frac{1}{2} [ \frac{1}{\bmth{q}^2-
\{ \epsl E_1(p)-\epsl' E_1(p') \}^2}
+\frac{1}{\bmth{q}^2- \{ \eta E_2(p)-\eta' E_2(p') \}^2} ],        \label{7}
 \end{eqnarray}
where $\bmth{q}= \bmth{p}-\bmth{p}'$. The retardation effects are
included in this equation.

\subsection{\it WFO of the $\,^1S_0$ state}

Let us next study some fundamental feature of the equation. When it is applied
 to the system in which the pair
 annihilation of the constituents can occur, an important physical
 quantity is the wave function at the origin(WFO). For example, the decay
 amplitude of the pseudo-scalar $Q\bar{q}$ meson via a weak boson is
proportional to the average WFO $\mbox{Tr}\{ \gamma_5\gamma_0\psi(0) \}$, where
$\psi(0)$
is the charge conjugated(with respect to the particle 2($\bar{q}$)) WFO. We
investigate, in Appendix, the asymptotic behaviour of the momentum-space wave
function by using the method given in Ref.\cite{Ito82}. We assume
instantaneous exchange of a gauge boson\footnote{The analysis in the Appendix
cannot be applied to the retarded interaction.}.
The average WFO thus obtained is finite. This result is
consistent with consideration on the covariant field theory. We note that the
average WFO becomes divergent in the limit of the one-body Dirac equation,
 for which we have the renormalization procedure\cite{Ito872}.

There are many "two-body Dirac equation" proposed. An interesting one from
 the point of view of the present paper is the one by Mandelzweig and
 Wallace\cite{Mandelzweig87}. Instead of redefining the two-body propagator,
they intended to include the effects of the higher-order interaction
(the crossed Feynman diagram)
 and obtained an equation which has the proper one-body limit and the
$E$-parity symmetry. An important difference
 from our equation is in the average WFO considered above. It is finite in
the Coulomb model but divergent if the transeverse part of the gauge-boson
 exchange is added\cite{Tiemeijer93}.

\subsection{\it On the three-body equation}

The next comment to be made is on the three-body equation in which the
interaction is assumed to be composed of
two-body interactions. It is almost straight forward to deduce the
three-body Hamiltonian.  However, we have to examine how to define the
projection factor in front of the quasipotential, since the sign of
the energy of a interacting sub-(two body)system depends on the
 reference frame. This shows that the applicability of the single-time
 formalism is more restrictive in the three-body system than in the
 two-body system. The effective range of the velocity of the subsystem relative
 to the three-body CM system cannot be large. Under this restriction,
we define the projection factor in the rest frame of the subsystem,
which is, then, transformed into the rest frame of the three-body system.
A part of the Hamiltonian which represent the interaction of the particle 1
 and 2, for example, becomes

\begin{equation}
   H_{12}=\sum_\epsl\epsl\Lambda_\epsl^1(\bmth{p}_1)
            V_{12}(\bmth{p}_1,\bmth{p}_2),                \label{31}
\end{equation}
when $m_1>m_2$.

Brown and Ravenhall pointed out that the three-body equation of some
category\footnote{They considered the two-body equation in an external
potential. The same argument is also applied to the three-body system
\cite{Sucher85}.}
 does not have proper solutions with the normalizable eigenfunction
\cite{Brown51}. This phenomenon is called "continuum dissolution(CD)".
The criteria for CD are as follows: (a) The equation decouples into the
 independent ones when a part of the interaction is switched off.
 (b) The subequations have solutions of the positive- and the negative-
energy continua. Our three-body equation does not fulfill (a) and is
free from CD.

\section{Equal-mass equation}

So far we have considered the unequal-mass equation. In this section we
investigate the equal-mass limit of it, for which the exchange symmetry is an
 issue.
We first note that the projection factor in front of the interaction term of
(\ref{4}) includes a part which violates this symmetry: It is shown that
\begin{equation}
  \Lambda^{(V)}=\Lambda_{+-}-\Lambda_{-+}         \label{e1}
\end{equation}
and the Heisenberg's exchange operator
\begin{equation}
  P_H=\frac{1}{4}(1+\bmth{\sigma}_1\cdot\bmth{\sigma}_2)
   (1+\bmth{\rho}_1\cdot\bmth{\rho}_2)P_M                    \label{e2}
\end{equation}
anticommute for $m_1=m_2$, where the operator $P_M$ exchanges the momenta
(or coordinates). $\Lambda^{(V)}$ violates the symmetry since the
 remaining part of the Hamiltonian is commutable with $P_H$.
For equal mass, we have two equations. One is the equation (\ref{4}) with
$m_1=m_2$ and another is obtained by assigning a minus sign in front of
$\Lambda^{(V)}$, which is the equal-mass limit of the equation with
$m_2>m_1$. It is the conjugate equation of (\ref{4}) in the sence that
(\ref{4}) is converted into it by the transformation $P_H$.
 It is easy to show that these equations have the common eigenvalue
spectrum: If an eigenfunction $\chi_n$ of (\ref{4}) belongs to some eigenvalue
$M_n$, $P_H\chi_n$ is the solution of the conjugate equation with the same
 eigenvalue. However, $\chi_n$ does not have the definite $P_H$-parity.

A way to recover the exchange symmetry is averaging the Hamiltonians
of the equation (\ref{4}) and the conjugate one. The resulting is the Salpeter
equation. The eigenvalue of this equation is different from the above $M_n$.
 The difference is, however, small since it is of the 4th order
in the symmetry-breaking part of the Hamiltonian.
The quarkonium phenomenology for the equal-mass constituents is given in
Ref.\cite{Ito90}.

 \section*{Acknowledgements}

The author would like to thank Professor G.A. Kozlov for useful communication
on the two-time Green's function.




\subsection*{Appendix }

We examine the asymptotic
($p\rightarrow\infty$) behavior of the momentum-space wave function and
 show that the average WFO $\mbox{Tr}\{ \gamma_5\gamma_0\psi(0) \}/\sqrt{2}$
 is finite.\footnote{See Ref.\cite{Ito88} and references therein, for
 the Salpeter equation.}


There are 4 partial amplitudes $h_{\epsl\eta}(p)$ in the ${}^1S_0$ state,
with which the wave function is expanded as

\begin{equation}
\chi(\bmth{p})=\sum_{\epsl\eta}\sum_{r}c_ru_\epsl^r(-\bmth{p})
v_\eta^{-r}(\bmth{p})h_{\epsl\eta}(p)( \frac{1}{16\pi E_1E_2} )^{1/2},
                                                      \label{pw}
\end{equation}
where $c_{1/2}=-c_{-1/2}=1/\sqrt{2}$ and the spinors $u$ and $v$, for the
particle 1 and 2 respectively, are defined in \cite{Ito82}.

The average WFO for the annihilation decay through the axial-vector current
 is given by

\begin{eqnarray}
\lefteqn{ \frac{1}{\sqrt{2}}\mbox{Tr}\{ \gamma_5\gamma_0\psi(0) \}
 = \frac{1}{\sqrt{8}\pi}\int (\frac{1}{E_1E_2})^{1/2} }     \nonumber
 \\ & & \times\sum_{\epsl\eta}
\{ \sqrt{(E_1+\epsl m_1) (E_2+\eta m_2)}-\epsl\eta\sqrt{ (E_1-\epsl m_1)
 (E_2-\eta m_2) } \}h_{\epsl\eta}(p)p^2dp. \hspace{6mm}           \label{wf}
\end{eqnarray}

We first assume the Coulomb potential. The partial-wave equation for the
 ${}^1S_0$ state is given by

\begin{eqnarray}
 \lefteqn{ \{ E-\epsl E_1(p)-\eta E_2(p) \}h_{\epsl\eta}(p)
=-\epsl\frac{\alpha}{4\pi}\sum_{\epsl'\eta'}\int dq } \nonumber \\
 & & \times \frac{q}{p}[ \frac{1}{E_1(p)E_1(q)E_2(p)E_2(q)} ]^{1/2}
 [\{ A^1_{\epsl\epsl'}A^2_{\eta\eta'}
+\epsl\epsl'\eta\eta'A^1_{-\epsl-\epsl'}A^2_{-\eta-\eta'} \}Q_0(z) \nonumber \\
 & &  \mbox{} +\{\epsl\epsl' A^1_{-\epsl-\epsl'}A^2_{\eta\eta'} +
 \eta\eta'A^1_{\epsl\epsl'}A^2_{-\eta-\eta'} \}Q_1(z)] h_{\epsl'\eta'}(q),
                       \label{pwe}
 \end{eqnarray}
where $z=(p^2+q^2)/2pq$ and $Q_\ell(z)$ is the Legendre's function.
$A^i_{\epsl\epsl'}$ is defined by

\[ A^i_{\epsl\epsl'}=\sqrt{(E_i(p)+\epsl m_i)(E_i(q)+\epsl' m_i)}. \]

 The asymptotic behavior of the wave function is determined from the integral
 region near the infinity.
 We then expand the both sides of (\ref{pwe}) into the series of $1/p$ and
 $1/q$. We assume the power behavior of the wave function for large $p$.
The independent amplitudes are chosen to be
$ h_A(p)\equiv\sum_\epsl h_{\epsl\epsl}(p)$,
$h_B(p)\equiv\sum_\epsl\epsl h_{\epsl\epsl}(p)$,
$h_C(p)\equiv\sum_\epsl h_{\epsl-\epsl}(p)$,
and
$h_D(p)\equiv\sum_\epsl\epsl h_{\epsl-\epsl}(p)$,
which are expanded, in the high-momentum region, in power series of $1/p$:
\[ h_X(p)=\sum_n C_X^np^{-\beta_X-2n-1}. \]

Integrals on the right-hand side can be done if we neglect infrared-divergent
terms which are irrelevant to the leading asymptotic behavior. Now, we can
determine the asymptotic indices $\beta_X$'s from
consistency\cite{Ito82}:
We get, for $h_A$ and $h_B$,

\begin{equation}
 2C_A^0 p^{-\beta_A}- EC_B^0 p^{-\beta_B-1}=
  \frac{\alpha}{\pi}C_A^0 \frac{\pi}{1-\beta_A}\cot(\frac{\pi}{2}\beta_A)
p^{-\beta_A}    \label{ia}
\end{equation}
and
\begin{equation}
 2C_B^0 p^{-\beta_B}- EC_A^0 p^{-\beta_A-1}=
  \frac{\alpha}{\pi}C_B^0
\frac{\pi(1-\beta_B)}{\beta_B(2-\beta_B)}\tan(\frac{\pi}{2}\beta_B)p^{-\beta_B},    \label{ib}
\end{equation}
where the terms of the higher power in $1/p$ are neglected.
If we neglect the second term in the left-hand sides of (\ref{ia}),
we find $\beta_A$ in the range $1<\beta_A<2$
\footnote{See Ref.\cite{Ito82} for the details.}
 and get $\beta_B=\beta_A+1$ from (\ref{ib}). We obtaine another series
by neglecting the second term in (\ref{ib}). For this,
 $\beta_B$ is found to be in the range $2<\beta_0<\beta_B<3$,
where the lower bound $\beta_0$ corresponds to the
upper bound $4/\pi$ of $\alpha$ above which the index $\beta_A$ from
(\ref{ia}) becomes complex.
$\beta_A$ of the second series is given by $\beta_A=\beta_B+1$.

The asymptotic amplitudes $h_C$ and $h_D$ are determined dependently on
$h_A$ and $h_B$.
We get, for the minimum indices

\begin{equation}
 \beta_C=\min(\beta_A+2,\: \beta_B+1)
\end{equation}
\begin{equation}
 \beta_D= \beta_A+1.
\end{equation}
We see that
the average WFO (\ref{wf}) is finite, because
\[  \beta_B>1 \: \mbox{ and }\: \beta_C>2               \]
hold for the asymptotic amplitudes.
This conclusion is valid even if the
 instantaneous exchange(transverse part) of the gauge boson is added.



\nocite{Feynman49}
\nocite{Hayashi52}
\nocite{Breit30}
\nocite{Ito901}
\nocite{Ito93}
\nocite{Aitchison82}
\bibliographystyle{unsrt}
\bibliography{quark}

\end{document}